\documentstyle[twocolumn]{mn}
\input{psfig.sty}
\title[Polarization Study of Radio Galaxies and Quasars]
{A Polarization Study of Radio Galaxies and Quasars selected from
the Molonglo Complete Sample}

\author[C.H. Ishwara-Chandra, D.J. Saikia, V.K. Kapahi \& McCarthy, P.J]
{C.H. Ishwara-Chandra$^{1,2}$, D.J. Saikia$^1$, V.K. Kapahi$^1$  and
P. J. McCarthy$^3$ \\
$^1$National Centre for Radio Astrophysics, TIFR, Post Bag 3, 
Ganeshkhind, Pune 411 007, India \\
$^2$ Joint Astronomy Program, Department of Physics, Indian 
Institute of Science, Bangalore, 560 012, India \\
$^3$The Observatories of the Carnegie Institution of Washington,
813 Santa Barbara st., Pasadena, CA 91101, USA \\
}
 
\date{}
\begin {document}
\maketitle
\begin{abstract}
We present total-intensity and linear-polarization observations made 
with the Very Large Array at $\lambda$20 and 6\,cm of a representative
sample of 42 radio galaxies and quasars selected from the Molonglo 
Complete Sample. The sources have been chosen to be of large 
size to probe the depolarizing medium on these scales 
using our present data and later with observations at lower 
frequencies with the Giant Metrewave Radio Telescope. The $\lambda$20 
and 6\,cm data are of similar resolutions and show that depolarization 
between these two wavelengths is seen largely in only those lobes which are
within about 300 kpc of the parent galaxy. Examination of the depolarization
of the lobes with arm-length asymmetry shows that depolarization is observed 
predominantly for the lobe which is closer to the nucleus. There is also 
a trend for the lobe
closer to the nucleus to be brighter, consistent with the scenario that the
nearer lobe is interacting with a denser environment which is responsible
for the higher depolarization and greater dissipation of energy.
We have also examined  the depolarization asymmetry of the lobes on 
opposite sides of the nucleus for galaxies and quasars. This shows that
the depolarization asymmetry for quasars is marginally higher than that for galaxies.
The depolarization properties of our sample are possibly determined by an 
asymmetric environment as well as the effects of orientation. 
\end{abstract}


\section{Introduction}
The discovery by Laing (1988) and Garrington et al. (1988) 
that double radio sources depolarize less rapidly on the side with 
the radio jet than on the opposite side provides the strongest 
evidence that the observed jet is on the approaching side, and its 
apparent asymmetry is due to relativistic beaming. The approaching 
side is seen through less of the depolarizing medium, and the 
Laing-Garrington effect can be understood as an orientation 
effect (cf. Garrington, Conway \& Leahy 1991; hereinafter referred to
as GCL91; Garrington \& Conway 1991). In addition, Liu \& 
Pooley (1991) found the more depolarized side of the source to have 
a steeper radio spectrum.  A detailed study of a sample of quasars 
by Dennett-Thorpe et al. (1997) shows that the spectrum of the 
high-brightness features is indeed flatter on the jet side, while 
the low-brightness features have a flatter spectrum on the side 
with the longer lobe. They suggest that this is due to relativistic 
bulk motion in the high brightness features and differential 
synchrotron ageing in the extended emission.

The correlation of depolarization asymmetry with jet 
sidedness is also relevant for testing the unified scheme in which 
quasars and BL Lac objects are believed to be inclined at small 
angles to the line-of-sight while radio galaxies lie close to the 
plane of the sky (Barthel 1989; Antonucci 1993; 
Urry \& Padovani 1995). In this scheme, 
the radio galaxies should exhibit a similar correlation of 
depolarization asymmetry with jet sidedness but of a smaller 
magnitude since the galaxies are inclined at large angles to the 
line of sight and the differential path length between the two 
lobes is smaller (cf. Holmes 1991; Saikia, Garrington \& Holmes 
1997). In addition to the effects of orientation, the 
depolarization of the lobes will also be affected by any asymmetry 
in the distribution of gas in the vicinity of the radio source. 
The possibility of an intrinsic asymmetry in the distribution of 
gas was suspected from the fact that the lobe on the jet side, 
which is approaching us, is often closer to the nucleus (Saikia 1981).
An intrinsic asymmetry was demonstrated clearly by McCarthy, 
van Breugel \& Kapahi (1991) who showed that there was invariably 
more emission-line gas on the side of the source which is closer 
to the nucleus. For a sample of 12 radio galaxies observed with 
the Very Large Array (VLA), Pedelty et al. (1989) found the 
arm-length ratios to be correlated with the amount of depolarization 
and emission line gas. For the radio galaxies which are at large 
angles to the line-of-sight, it appears that the correlation of 
depolarization with arm-length asymmetry is stronger than with 
jet asymmetry, while the reverse is true for quasars (cf. Laing 1993). 

In this paper we have studied the depolarization properties 
of a  matched sample of radio galaxies and quasars 
selected from the Molonglo Reference Catalogue, to examine 
the effects of their environment as well as orientation  on 
structural and depolarization asymmetries of these objects. Our 
objects were chosen to be of large angular and linear size to 
probe the depolarizing medium on these scales using high-frequency 
observations with the VLA and lower-frequency observations with 
the Giant Metrewave Radio Telescope (GMRT) in the future. We have 
also started a programme to make optical narrow-band images to 
study the emission-line gas as well as broad-band continuum images 
to study the environments of these objects. In this paper we 
present our total-intensity and linear polarization observations 
with the VLA BnA and CnB arrays at the L- and C-bands respectively. 
In Section 2 we describe the sample of sources, while the observations 
and observational results are presented in Sections 3 and 4 
respectively. We then present our estimates of the depolarization 
of the lobes  and hotspots, discuss their relationship with linear size and 
arm-length asymmetry, and any difference in depolarization 
asymmetry between radio galaxies and quasars. A study of the RM 
distributions in these sources and any relationship with 
line-emitting gas will be presented in a later paper.

\begin{table} \caption{ The sample of sources }
\begin{tabular}{cclrcrr }
Source & Id   & z & S$_{1365}$ & P$_{1365}$ & LAS & LLS \\
 Name  &      &      & mJy  & W/Hz/sr   & $''$& kpc \\
           &   &        &        &         &         &     \\
0017-207 & Q & 0.545 &  467 &  25.76 &  98   &  722 \\
0058-229 & Q & 0.706 &  396 &  25.95 &  63   &  505 \\
0133-266 & Q & 1.53  &  348 &  26.61 &  53   &  455 \\
0137-263 & G & 1.1   &  509 &  26.51 &  79   &  679 \\
0148-297 & G & 0.41  & 2778 &  26.28 & 148   &  961 \\
           &   &        &        &         &         &     \\
0325-260 & G & 0.638 &  286 &  25.70 &  58   &  447 \\
0346-297 & G & 0.413 &  620 &  25.65 & 142   &  929 \\
0428-281 & G & 0.65  &  956 &  26.25 &  79   &  617 \\
0437-244 & Q & 0.84  &  459 &  26.19 & 128   & 1059 \\
0454-220 & Q & 0.533 & 1993 &  26.35 &  94   &  683 \\
           &   &        &        &         &         &     \\
0551-226 & G & 0.8   &  330 &  26.00 &  53   &  438 \\
0937-250 & G & 0.9   &  445 &  26.28 &  72   &  607 \\
0938-205 & G & 0.371 &  444 &  25.39 &  72   &  445 \\
0947-249 & G & 0.854 & 1487 &  26.75 &  71   &  595 \\
0955-283 & G & 0.8   &  470 &  26.14 &  96   &  791 \\
           &   &        &        &         &         &     \\
1022-250 & G & 0.34  &  338 &  25.20 &  61   &  363 \\
1023-226 & G & 0.586 &  298 &  25.64 &  66   &  495 \\
1025-229 & Q & 0.309 &  489 &  25.28 & 198   & 1105 \\
1026-202 & G & 0.566 &  664 &  25.95 &  61   &  456 \\
1029-233 & G & 0.611 &  403 &  25.80 &  80   &  609 \\
           &   &        &        &         &         &     \\
1052-272 & Q & 1.103 &  553 &  26.56 &  89   &  760 \\
1107-218 & G & 1.5   &  302 &  26.53 &  62   &  531 \\
1107-227 & G & 2.0   &  781 &  27.40 &  72   &  590 \\
1126-290 & G & 0.41  & 1086 &  25.89 & 111   &  725 \\
1224-208 & G & 1.5   &  268 &  26.57 &  61   &  519 \\
           &   &        &        &         &         &     \\
1226-297 & Q & 0.749 &  435 &  26.00 &  74   &  597 \\
1232-249 & Q & 0.352 & 1988 &  25.97 & 110   &  660 \\
1247-290 & Q & 0.77  &  696 &  26.27 &  60   &  489 \\
1257-230 & Q & 1.109 &  788 &  26.65 &  53   &  458 \\
1358-214 & G & 0.5   &  317 &  25.34 &  96   &  681 \\
           &   &        &        &         &         &     \\
2035-203 & Q & 0.516 &  752 &  25.87 &  71   &  514 \\
2040-236 & Q & 0.704 &  428 &  25.89 &  64   &  510 \\
2042-293 & G & 1.9   &  414 &  26.99 &  79   &  655 \\
2045-245 & G & 0.73  &  639 &  26.21 &  76   &  611 \\
2118-266 & G & 0.343 &  403 &  25.24 &  89   &  526 \\
           &   &        &        &         &         &     \\
2132-235 & G & 0.81  &  333 &  26.02 &  58   &  482 \\
2137-279 & G & 0.64  &  455 &  25.93 &  57   &  442 \\
2213-283 & Q & 0.946 &  820 &  26.53 &  75   &  633 \\
2311-222 & G & 0.434 &  965 &  25.85 &  92   &  616 \\
2325-213 & G & 0.58  &  986 &  26.17 &  84   &  632 \\
           &   &        &        &         &         &     \\
2338-290 & Q & 0.446 &  427 &  25.51 &  79   &  532 \\
2348-235 & G & 0.952 &  503 &  26.34 &  67   &  570 \\
\end{tabular}
\end{table}

\section{The sample of sources}
Our sample has been selected from a complete sample of sources, 
the MRC/1Jy sample (McCarthy et al. 1996; Kapahi et al. 1998a, 1998b) 
which is a subset of the Molonglo Reference Catalogue (Large et 
al. 1981).  The MRC/1 Jy sample consists of 558 sources which 
satsify the following criteria: $S_{408} \geq 0.95 $ Jy; lie in 
the declination range $-30^{\circ} < \delta < -20^{\circ}$ and 
the right asension ranges 20$^h$ 20$^m$ to 06$^h$ 15$^m$ and 
09$^h$ 25$^m$ to 14$^h$ 03$^m$ in B1950 co-ordinates; and lie
outside the Galactic plane ($| b | > 20^{\circ}$). We have 
chosen FR II (Fanaroff \& Riley 1974) 
sources from this sample with an angular size 
larger than about an arcminute to probe the depolarizing medium 
on these scales using the VLA and GMRT. We have chosen all quasars
with a measured redshift greater than about 0.3, and a matched
sample of galaxies of similar luminosity and redshift. 
The final sample of sources consists of 27 radio galaxies and 15 
quasars, which are listed in Table 1. The redshifts for 
0137$-$263, 0551$-$226, 0937$-$250, 0955$-$283, 1107$-$218,
1107$-$227, 1224$-$208 and 2042$-$293 have been estimated from
their K magnitudes and are listed to only the first decimal place.
The flux density at
1365\,MHz and the spectral index between 1365 and 4935\,MHz used for
calculating the luminosity have been estimated from our observations.
The luminosity, P, is in units of WHz$^{-1}{\rm sr}^{-1}$ 
and log P$_{1365}$ is listed in the Table. 
The largest angular size
(LAS) is expressed in arcsec and the corresponding largest
linear size (LLS) is in kpc. The LAS has been estimated from the
1365\,MHz images and represents the separation between the high 
brightness peaks in the outermost regions of emission on opposite
sides of the nucleus. 
The radio galaxies and quasars in our sample have similar 
distributions of redshift, luminosity and linear size. For the 
galaxies, the median values of redshift and luminosity are  0.64 
and 26.02 WHz$^{-1}{\rm sr}^{-1}$ , compared to 0.71 and 
26.00 WHz$^{-1}{\rm sr}^{-1}$ for the quasars. The 
corresponding values of angular and linear sizes are  72$''$ and 
595 kpc for the galaxies and 75 $''$ and 597 kpc for the quasars. 
We have assumed an
Einstein de-Sitter Universe with H$_{\circ}$ = 50 km s$^{-1}$ Mpc$^{-1}$.

\section{Observations and analyses}
The observations were made with scaled arrays of the 
Very Large Array of National radio Astronomy Observatory
(Thompson 1980) in order to get similar resolutions at the L- and 
C-bands. A summary of the observations is presented in Table 2. 
Short scans of unresolved sources selected from the VLA Calibrator Manual 
(Perley 1996) were interspersed for phase calibraton. 
At both the bands two widely separated IFs with bandwidths of 
50\,MHz were used except for the one centered at 1665\,MHz where the
bandwidth was only 25\,MHz. The IFs were separated by 
about 300\,MHz so that the 
rotation measure (RM) could be determined reliably.  
One source, 1358$-$214, was not observed at the C band due 
to technical problems.   Most sources were observed in two scans at different 
hour angles, each lasting about 10 min. 

\begin{table} \caption{ Observing schedule }
\begin{tabular}{l l l l l}

Array    &Freq. of & Band-  & Date  &Total  \\
Conf.    &obs.     & width  &       &time  \\
         & MHz     & MHz    &       & hr   \\
         &           &        &     &   \\
BnA      &1365   &  50 & 1995 Sept. 20 & 16.0   \\
         &1665   &  25 &            &           \\
         &           & &       &        \\
CnB      &4635   &  50 & 1996 Jan. 20,31& 10.5  \\
         &4935   &  50 &              &           
\end{tabular}
\end{table}

\begin{table} \caption{ Image parameters }

\begin{tabular}{lrrrrrrr}
Source & \multicolumn{3}{c}{beam } & \multicolumn{2}{c}{22\,cm} & \multicolumn{2}{c}{6.08\,cm} \\
Name   &maj & min &PA &$\sigma _I$ &$\sigma _p$ & $\sigma _I$ &$\sigma _p$ \\
 & \multicolumn{2}{c}{$''$}
&\multicolumn{1}{c}{$^\circ$}& \multicolumn{4}{c}{$\mu$Jy/beam} \\
           &     &      &      &       &    &    &     \\
0017$-$207 & 4.5 & 3.5  & $+$70& 180   & 60 &  55& 55  \\
0058$-$229 & 4.2 & 3.5  & $+$80& 107   & 65 &  44& 50  \\
0133$-$266 & 5.0 & 3.0  & $+$60&  86   & 60 &  43& 65  \\
0137$-$263 & 5.0 & 3.0  & $+$60& 149   & 60 &  55& 55  \\
0148$-$297 & 4.5 & 3.5  & $+$60& 246   & 70 & 211& 50  \\
           &     &      &      &       &    &    &     \\
0325$-$260 & 4.2 & 4.2  &   00 & 119   & 65 &  40& 50  \\
0346$-$297 & 4.2 & 4.2  &   00 & 141   & 60 &  55& 50  \\
0428$-$281 & 4.5 & 3.5  & $-$70& 146   & 65 &  50& 65  \\
0437$-$244 & 4.5 & 3.2  & $-$70&  96   & 65 &  50& 55  \\
0454$-$220 & 4.5 & 3.0  & $-$70& 194   & 60 &  75& 60  \\
           &     &      &      &       &    &    &     \\
0551$-$226 & 6.0 & 3.5  & $-$50& 111   & 60 &  50& 50  \\
0937$-$250 & 8.0 & 4.5  & $-$50& 142   & 61 &  55& 55  \\
0938$-$205 & 7.5 & 5.0  & $-$50& 122   & 58 &  54& 60  \\
0947$-$249 & 8.0 & 4.0  & $-$50& 178   & 65 & 109& 50  \\
0955$-$283 & 6.0 & 6.0  &   00 & 180   & 66 &  65& 55  \\
           &     &      &      &       &    &    &     \\
1022$-$250 & 6.0 & 6.0  &    00& 238   & 60 &  67& 65  \\
1023$-$226 & 8.0 & 4.5  & $-$50& 148   & 58 &  47& 50  \\
1025$-$229 & 8.0 & 4.5  & $-$50& 167   & 62 &  70& 50  \\
1026$-$202 & 8.0 & 4.2  & $-$50& 161   & 60 &  56& 65  \\
1029$-$233 & 6.0 & 4.5  & $-$50& 204   & 58 &  72& 55  \\
           &     &      &      &       &    &    &     \\
1052$-$272 & 6.0 & 5.0  & $-$50& 150   & 60 &  50& 60  \\
1107$-$218 & 6.5 & 4.5  & $-$50& 125   & 60 &  50& 55  \\
1107$-$227 & 6.5 & 4.5  & $-$50& 234   & 65 &  60& 60  \\
1126$-$290 & 6.0 & 5.0  & $-$50& 201   & 60 &  81& 55  \\
1224$-$208 & 7.0 & 4.2  & $-$50& 192   & 60 &  75& 60  \\
           &     &      &      &       &    &    &     \\
1226$-$297 & 7.0 & 4.5  & $-$50& 361   & 61 &  82& 55  \\
1232$-$249 & 7.0 & 4.5  & $-$50&1352   & 85 &  87& 55  \\
1247$-$290 & 7.0 & 5.5  & $-$50& 250   & 60 &  75& 60  \\
1257$-$230 & 7.0 & 4.5  & $-$50& 545   & 70 &  53& 50  \\
1358$-$214 & 6.5 & 3.0  & $-$50& 178   & 60 & $-$& $-$ \\
           &     &      &      &       &    &    &     \\
2035$-$203 & 7.0 & 3.0  & $+$50& 238   & 60 &  79& 65  \\
2040$-$236 & 6.4 & 3.0  & $+$50& 194   & 65 &  59& 50  \\
2042$-$293 & 8.4 & 3.0  & $+$40& 192   & 65 &  65& 60  \\
2045$-$245 & 7.5 & 3.0  & $+$50& 155   & 60 &  53& 65  \\
2118$-$266 & 7.5 & 3.0  & $+$50& 244   & 60 &  72& 55  \\
           &     &      &      &       &    &    &     \\
2132$-$236 & 7.5 & 3.0  & $+$50& 100   & 70 &  51& 60  \\
2137$-$279 & 7.5 & 3.0  & $+$50& 197   & 60 &  51& 60  \\
2213$-$283 & 6.5 & 3.0  & $+$50& 166   & 65 &  55& 55  \\
2311$-$222 & 5.0 & 3.0  & $+$60& 201   & 60 &  62& 65  \\
2325$-$213 & 5.0 & 3.0  & $+$60& 159   &148 &  55& 50  \\
           &     &      &      &       &    &    &     \\
2338$-$290 & 5.0 & 3.5  & $+$50& 261   & 60 &  76& 55  \\
2348$-$235 & 5.0 & 3.0  & $+$60& 113   &136 &  76& 60  \\

\end{tabular}
\end{table}

The data for each IF at both bands were edited and 
calibrated separately using the NRAO {\tt AIPS}
package. All flux densities are on the Baars et al. (1977) scale,
with 3C48 and 3C286 being the primary flux density calibrators. 
To determine the absolute position angle both 3C138 and
3C286 were observed while
the instrumental polarization was determined by observing the 
unresolved calibtrators, over a range of parallactic angles. 
We have corrected for ionospheric Faraday 
rotation, which could be significant at the L-band, using the routines in 
the {\tt AIPS} package (cf. Chiu 1975). Comparing the polarization 
position angles of 3C286 and 3C138 before and after applying the ionospheric
Faraday correction based on this model, we find the difference in PA to be
within about 5$^\circ$. 
After completing the continuum and polarization calibration, 
images of the sources  were made with uniform weighting to all baselines using 
the {\tt AIPS} task {\tt MX} and {\tt IMAGR}. Several iterations of 
self calibrations were also done to correct for residual phase errors.
The final data set was used to make images in the Stokes parameters 
I, Q and U.  A given source
has been restored with a beam of the same size at both the L and C bands. 
We have also made circular polarization or Stokes V images of all 
sources to estimate the rms noise in our maps. The ratio of the peak
intensity to the rms noise in the total-intensity images is typically 
more than about 1000:1 at the L band.  

The maps of polarized intensity, $p = (Q^2 + U^2)^{1/2}$, and 
position angle, $\chi = 0.5 {\rm tan}^{-1}(U/Q)$ were made 
by combining the Q and U maps.  The resulting polarized intensity 
image has a positive 
bias which has been corrected using the {\tt AIPS} task {\tt POLCO}, 
and also all pixels of amplitude $\leq$ 2$\sigma$ were blanked, 
where $\sigma$ is the noise in the Q or U maps. The polarized flux 
density was estimated from this image, while the rms noise on the 
polarized intensity map has been estimated before applying the bias correction. 
The images of the sources at one IF in each of the two bands are 
presented in Figure 1, with the fractional polarization vectors 
superimposed on the total-intensity contours. The image of the 
source, 1358$-$214,
for which we have data at only the L-band is shown at the end
of Figure 1. The size of the 
restoring beam and the rms noise in the total-intensity and 
polarization images for these frequencies are listed in
Table 3. The corresponding values
for the other frequency in each band are similar.

\subsection{Observational results}
We have estimated the total intensity and polarized flux density of each 
lobe of a source by specifying a rectangular box such that all the 
regions of the lobe with total flux density more than  about 5 times
the rms noise per beam in the total-intensity image
are included in the box.  Following Garrington et al. (1991), we have 
set the box based on the 20\,cm image and a box of the same size has 
been used for the $\lambda$6\,cm image. These values were used to 
estimate the degree of polarization and depolarization between these 
two wavelengths for the lobes. In addition, we have also estimated 
the degree of polarization and depolarization for the hotspots by 
specifying a small box of 5$\times$5 arcsec$^{2}$ centred on the 
pixel of maximum brightness in each lobe as seen on the $\lambda$20\,cm 
image.  In our images the pixel size is one arcsec at both bands. 
The integrated and peak values are listed in Table 4 which 
is arranged as follows. 
Column 1: source name; column 2: component designation where N, S, W and
E denote north, south, west and east respectively; columns 3 and 4: 
the total intensity, $S_I$, and the degree of scalar polarization, 
$m_l$=$(\Sigma p/\Sigma I)\times 100$ \%, for the entire 
component at $\lambda$22\,cm;  columns 5 and 6: the peak brightness 
of the lobe, $S_p$, and the degree of polarization of the hotspot, 
$m_{hs}$, at $\lambda$22\,cm; columns 7 to 10: same as columns 3 to 6, 
but at $\lambda$18\,cm; columns 11 to 14:  same as columns 3 to 6, 
but at $\lambda$6.47\,cm; columns 15 to 18: same as columns 3 to 6, 
but at $\lambda$6.08\,cm. 

\subsection{ Error estimates} 
For our interpretation we need to estimate the error in the degree 
of polarization and the depolarization ratio between the L and C bands. 
The measurement errors in the total- and polarized- flux density are 
estimated from the off-source rms fluctuations. In the region over 
which we measure the total flux density, this error is 
$N_I^{1/2} \delta _I $,  and the error in polarization maps is 
$N_p^{1/2} \delta _p $, where $N_I$ and $N_p$ are the number of 
non-blank beam areas and $\delta _I $ and $\delta _p $ are the 
off-source rms fluctuations per beam in the total- and polarized-intensity 
maps.  The fractional 
polarization $m$ and the fractional error in $m$ are 
$m = {\Sigma p \over \Sigma I} \times 100\% ~$ and 
$~ \delta m  = ({\sigma_p \over \Sigma p}  + 
{\sigma_I \over \Sigma I}) m $ respectively.  From the internal consistency 
of the data, we have estimated the uncertainity in polarization 
calibration to be better than about 0.3\%. Combining this with  
the fractional error in $m$, the resultant uncertainity in $m$ is  
$\sigma_m = [(\delta m)^2 + 0.3^2]^{1/2} \%.$ 

The depolarization DP and the error in DP are as follows:
DP$ = m_{20} / m_6 \ ; \
\sigma_{DP}  =  {\rm DP} \times [({\sigma_{m_{20}} \over m_{20}})^2  + 
({\sigma_{m_{6}} \over m_6})^2 ]^{1/2}$, 
since the errors on $m_{20}$ and $m_6 $ are uncorrelated.  
The errors in our estimates of fractional polarization are in 
the range 0.3\% - 0.6\%, while the estimated errors in DP are less 
than about 0.08 in most cases.

\section{Results and discussion}
In our sample of 41 sources with information at both bands,
there are a total of 82 lobes. However, 6 of these lobes,
namely 0346$-$297N, 0938$-$205N, 1025$-$229S, 1226$-$297N,
2040$-$236W and 2118$-$266W, have been excluded from the
analysis becasue they are very weak and diffuse and
their polarization information could not be determined reliably.
The derived parametes for the remaining 76 lobes are presented in Table 5
which is arranged as follows. Column 1: source name and letter
designation identifying the component; column 2: the linear
separation in kpc of the peak of the lobe, including the hotspot
at the angular resolution of the image,  from the radio core
or the optical position, if a core has not been detected; 
column 3:
the ratio of separation of the component from the core/optical
position  to that of the component on the opposite side; columns
4 and 5: the depolarization DP=m$_{20}$/m$_6$ and the error in
DP for the entire lobe including the hotspot; columns 6 and 7:
the depolarization and the error in DP for the hotspots.

The distributions of the degree of polarization of the lobes, m$_l$,
at $\lambda$6\,cm range from about 2.4 to 18.2\%, with a median
value of about 10\%. This is similar to the sample of sources
observed by GCL91, where the median value on the jet as well as
counter-jet side is 10.7\%. We have also
examined the distributions for the sample of 40 strong lobes defined
to be those with
S$_{peak}\geq$25 mJy at $\lambda$6\,cm, and find their polarization
to be similar to those of the weaker sources. The degree of polarization
for the hotspots in the samples of strong and weak lobes are also similar.
At $\lambda$20\,cm, the distribution of the degree
of polarization, m$_{20}$, ranges from 2.8  to 18.4\%, with a
median value of again about 10\%. The median values 
for the strong and weak source samples are about 9.3 and 11.0\,%
respectively. The weak sources have higher errors and the difference
is margnial. There is a similar difference  for the
hotspot values. The degree of polarization in our lobes is only marginally
higher than 
the jet sides at $\lambda$20\,cm of the GCL91 sample where the median
value is 8\%, although the
counter-jet sides in GCL91 exhibit significant depolarization. The values
of the depolarization parameter, DP = m$_{20}$/m$_6$, for the lobes range from
about 0.54 to 2.95, with median values of about 1 for the strong
lobes, and about 1.1 for the weaker ones. The values of DP for the hotspots are
similar.

\begin{figure*}
\vbox{
\vspace{-0.8 in}
\hbox{
\hspace{-0.60 in}
\psfig{figure=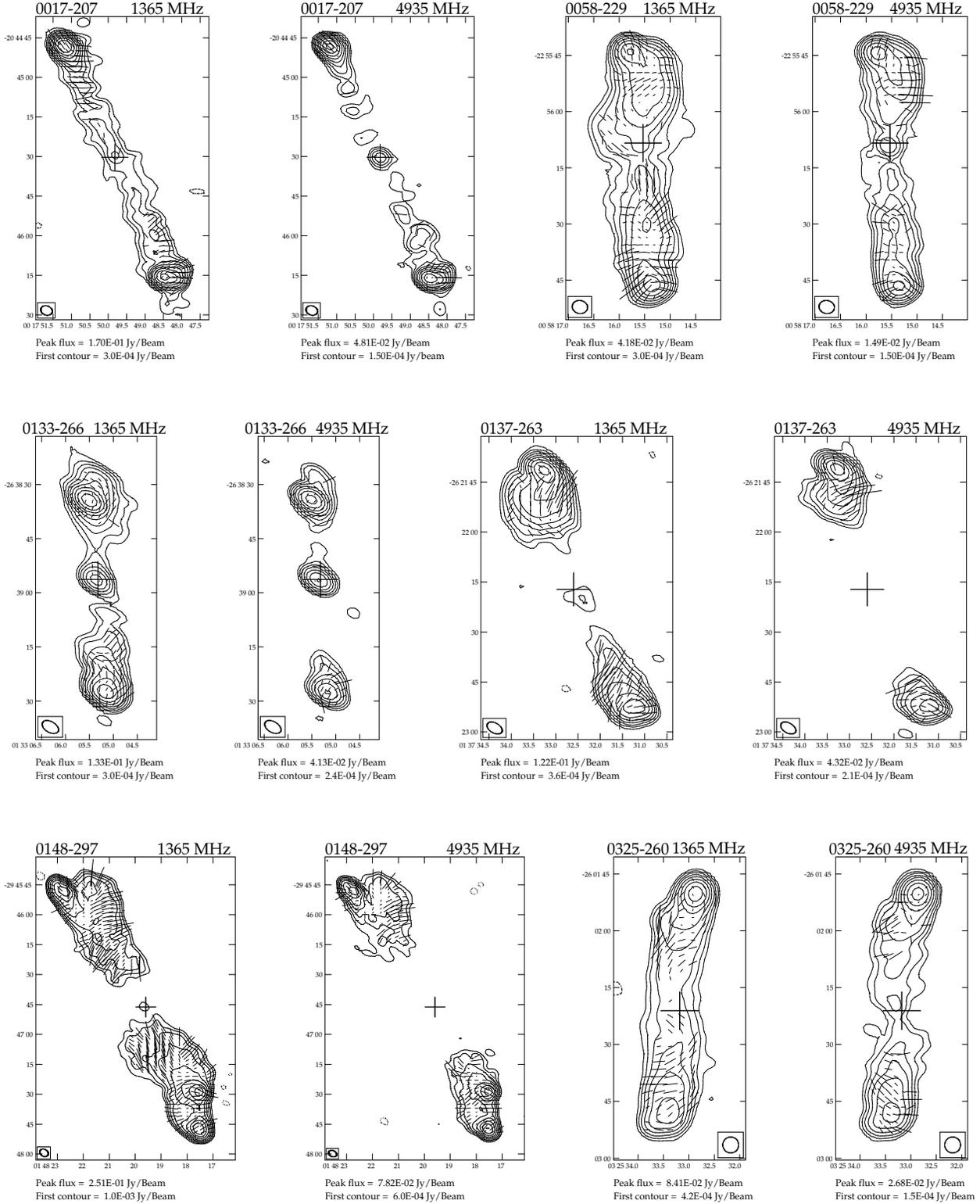}
}
\vspace{-1.5in}
}
\caption{Images of the sample of sources at 1365 and 4935\,MHz. The image for 1358$-$214, for which there is
data at only the L-band is shown at the end of this figure. The polarization vectors are superimposed on
the total-intensity contours. The restoring beam is shown by the ellipse enclosed within a box, while the
cross indicates the position of the optical object. The contour levels are 
-2, -1, 1, 2, 4, 8, 16, 32, 64, 128, 256, 512 times the base level given below each figure. One arc 
second length of polarization vector corresponds to 5\% polarization.  X axis is right ascension and Y axis is declination in B1950 co ordinates.}
\end{figure*}

\begin{figure*}
\vbox{
\vspace{1.5in}
\hbox{
\hspace{1.5in}
\psfig{figure=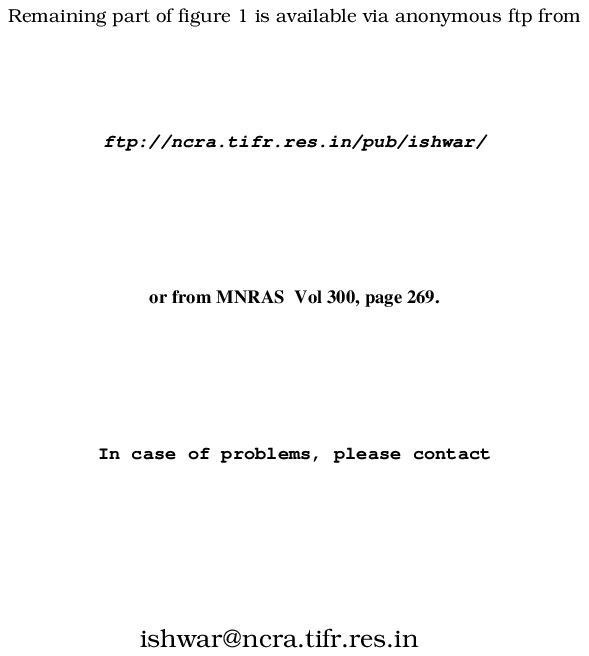,height=6.0in,width=4.0in}
}
\vspace{1.5in}
}
\end{figure*}


\begin{table*} \caption {Flux density and degree of polarization at different wavelengths}
\begin{tabular}{ll rrrr rrrr rrrr rrrr}
\multicolumn{2}{c}{Source} & \multicolumn{4}{c}{22\,cm }
				& \multicolumn{4}{c}{18\,cm} & \multicolumn{4}{c}{6.47\,cm}& \multicolumn{4}{c}{6.08\,cm} \\
Name      &Cp&$S_I$ & m$_l$    &S$_p$&m$_{h}$&$S_I$ & m$_l$ &S$_p$&m$_{h}$& $S_I$& m$_l$ &S$_p$ & m$_{h}$& $S_I$& m$_l$   &S$_p$ & m$_{h}$\\
          &    &      &      &     &     &      &      &     &     &      &     &      &      &      &     &      &      \\
0017$-$207&N  &260   &13.8  &170  &15.6 &191   &14.3  &140  &16.1 &75    &13.4 &51    &16.0  &70    &14.1 &48    &17.2  \\
          &S  &202   &9.4   &99   &8.6  &173   &8.9   &85   &8.3  &62    &9.7  &33    &6.6   &58    &7.6  &32    &6.7   \\
0058$-$229&N  &225   &5.8   &42   &10.3 &173   &7.6   &34   &13.1 &64    &8.9  &17    &12.5  &57    &8.5  &15    &12.2  \\
          &S  &158   &11.0  &30   &14.3 &125   &9.3   &25   &12.6 &49    &9.5  &13    &11.1  &44    &8.5  &12    &10.4  \\
0133$-$266&N  &196   &4.2   &133  &4.2  &155   &4.2   &110  &4.2  &58    &3.0  &44    &3.2   &53    &4.5  &41    &3.7   \\
          &S  &134   &8.3   &67   &6.4  &102   &7.4   &54   &6.8  &30    &6.9  &18    &6.3   &26    &7.7  &17    &7.5   \\
0137$-$263&N  &384   &12.5  &122  &8.3  &307   &12.0  &105  &8.3  &102   &12.6 &45    &11.0  &93    &13.5 &43    &11.2  \\
          &S  &125   &16.6  &37   &17.6 &99    &16.4  &30   &18.3 &31    &16.9 &12    &14.8  &28    &15.3 &11    &18.9  \\
0148$-$297&N  &1150  &17.3  &197  &24.2 &862   &18.7  &168  &26.0 &334   &15.0 &74    &24.0  &311   &16.6 &65    &27.0  \\
          &S  &1628  &18.3  &251  &21.3 &1289  &19.3  &217  &22.7 &460   &16.3 &92    &19.8  &436   &17.6 &78    &23.2  \\
          &   &      &      &     &     &      &      &     &     &      &     &      &      &      &     &      &      \\
0325$-$260&N  &168   &6.3   &84   &3.7  &140   &5.6   &70   &3.2  &53    &4.0  &29    &2.1   &48    &3.9  &27    &2.0   \\
          &S  &117   &9.9   &24   & 9.8 &93    &9.6   &19   &10.8 &35    &10.0 &8.1   &7.8   &30    &8.2  &7.1   &8.2   \\
0346$-$297&N  &71    &21.0  &3.9  &33.0 &45    &22.9  &2.8  &37.7 &20    &$-$  &1.2   & $-$  &14    & $-$ &0.97  & $-$  \\
          &S  &549   &18.4  &55   &17.1 &424   &17.7  &43   &16.6 &142   &15.3 &18    &13.5  &128   &16.1 &16    &14.5  \\
0428$-$281&E  &424   &9.4   &229  &4.9  &326   &9.0   &189  &5.0  &114   &6.9  &70    &4.4   &106   &7.5  &66    &4.6   \\
          &W  &532   &10.0  &318  &7.3  &426   &9.8   &271  &7.4  &154   &8.6  &105   &7.2   &142   &10.4 &98    &7.9   \\
0437$-$244&N  &280   &8.0   &66   &12.9 &197   &7.8   &53   &12.0 &73    &8.2  &22    &13.0  &66    &7.8  &20    &13.0  \\
          &S  &160   &12.2  &44   &6.8  &108   &9.2   &36   &7.0  &41    &10.3 &15    &5.7   &37    &6.7  &14    &4.2   \\
0454$-$220&N  &716   &6.9   &199  &7.6  &553   &8.6   &163  &9.0  &226   &10.5 &65    &9.0   &212   &10.6 &60    &9.6   \\
          &S  &1006  &10.5  &380  &10.9 &792   &10.5  &312  &11.4 &283   &10.9 &107   &11.7  &264   &10.7 &99    &11.8  \\
          &   &      &      &     &     &      &      &     &     &      &     &      &      &      &     &      &      \\
0551$-$226&N  &151   &9.6   &25   &14.1 &87    &12.7  &17   &15.9 &46    &9.2  &8.8   &14.8  &40    &11.8 &8.0   &14.5  \\
          &S  &179   &8.7   &29   &14.6 &104   &10.7  &19   &18.0 &45    &9.6  &9.1   &17.0  &43    &9.9  &8.3   &18.1  \\
0937$-$250&N  &175   &8.8   &56   &9.1  &129   &11.0  &46   &9.3  &31    &7.4  &17    &4.1   &30    &10.7 &17    &5.4   \\
          &S  &270   &5.8   &100  &5.8  &215   &5.0   &81   &4.7  &70    &5.2  &29    &4.3   &64    &5.4  &27    &5.2   \\
0938$-$205&N  &224   &2.3   &23   &7.2  &170   &2.9   &19   &7.6  &68    &$-$  &8.7   &$-$   &60    &$-$  &8     & $-$  \\
          &S  &220   &11.9  &40   &9.8  &167   &12.0  &32   &9.1  &66    &7.0  &16    &5.7   &59    &11.2 &16    &8.1   \\
0947$-$249&N  &1087  &10.4  &602  &13.1 &845   &12.3  &487  &13.4 &243   &15.1 &159   &13.0  &227   &16.2 &149   &14.0  \\
          &S  &336   &6.6   &223  &6.1  &267   &6.4   &180  &5.7  &89    &6.4  &64    &5.7   &83    &6.5  &60    &5.8   \\
0955$-$283&E  &171   &13.9  &53   &14.7 &122   &13.5  &43   &16.6 &44    &15.2 &17    &18.0  &40    &14.0 &17    &16.6  \\
          &W  &299   &8.7   &192  &9.5  &236   &8.7   &158  &9.7  &92    &8.3  &60    &9.2   &86    &8.6  &56    &9.4   \\
          &   &      &      &     &     &      &      &     &     &      &     &      &      &      &     &      &      \\
1022$-$250&E  &127   &5.6   &37   &8.7  &87    &5.1   &28   &8.2  &33    &4.5  &12    & 8.6  &29    &6.4  &11    & 7.2  \\
          &W  &212   &12.6  &52   &17.2 &160   &12.7  &41   &17.1 &62    &12.4 &19    &13.8  &57    &12.2 &18    &13.1  \\
1023$-$226&N  &197   &5.2   &83   & 6.9 &156   &4.7   &68   &6.2  &64    &3.7  &30    &4.4   &60    &4.1  &29    &5.7   \\
          &S  &75    &2.8   &46   & 2.9 &60    &2.8   &37   &3.6  &25    &2.4  &16    &2.5   &22    &2.5  &15    &3.1   \\
1025$-$229&N  &250   &9.0   &78   &5.8  &181   &8.8   &67   &5.9  &59    &7.1  &27    &5.8   &52    &8.5  &24    &5.9   \\
          &S  &176   &10.7  &20   &9.4  &122   &10.5  &17   &12.4 &38    &$-$  &6.6   & $-$  &33    &$-$  &6.2   & $-$  \\
1026$-$202&N  &262   &11.4  &69   &9.8  &215   &11.1  &57   &9.5  &81    &9.3  &24    &10.8  &75    &9.6  &22    &11.1  \\
          &S  &401   &6.8   &134  &8.3  &321   &7.8   &110  &9.8  &117   &8.0  &45    &7.9   &111   &7.7  &43    &7.9   \\
1029$-$233&E  &290   &14.2  &116  &11.4 &220   &15.1  &94   &12.7 &91    &14.4 &40    &13.9  &87    &14.3 &38    &14.3  \\
          &W  &86    &11.0  &38   &9.4  &66    &10.3  &30   &9.9  &28    &9.9  &13    &9.2   &25    &10.7 &12    &10.8  \\
          &   &      &      &     &     &      &      &     &     &      &     &      &      &      &     &      &      \\
1052$-$272&N  &345   &6.2   &192  &5.7  &263   &7.1   &153  &6.4  &74    &8.7  &49    &8.1   &70    &8.4  &46    &8.0   \\
          &S  &205   &7.9   &69   &10.7 &164   &7.6   &56   &9.6  &54    &7.8  &20    &8.6   &51    &6.7  &20    &8.1   \\
1107$-$218&E  &105   &5.1   &56   &4.1  &83    &6.9   &48   &5.0  &33    &5.3  &21    &4.2   &30    &5.8  &20    &4.2   \\
          &W  &197   &7.2   &124  &10.5 &153   &7.8   &103  &11.3 &60    &8.0  &43    &10.7  &55    &10.0 &40    &12.0  \\
1107$-$227&N  &520   &8.0   &346  &5.1  &398   &7.5   &271  &5.4  &104   &7.4  &81    &8.4   &96    &7.7  &75    &9.1   \\
          &S  &261   &12.6  &153  &15.7 &195   &11.6  &114  &15.1 &47    &9.2  &30    &8.7   &41    &9.6  &24    &8.1   \\
1126$-$290&N  &352   &11.6  &73   &6.5  &275   &11.3  &60   &6.8  &71    &14.4 &24    &5.8   &65    &13.7 &23    &5.8   \\
          &S  &734   &12.1  &169  &13.1 &565   &11.9  &146  &13.6 &193   &12.4 &64    &13.4  &176   &12.2 &62    &13.2  \\
1224$-$208&N  &188   &15.4  &71   &19.5 &141   &18.1  &54   &20.6 &42    &19.5 &20    &16.9  &38    &18.2 &18    &18.1  \\
          &S  &80    &10.0  &34   &10.6 &56    &11.4  &25   &10.9 &20    &5.6  &9.6   &9.2   &18    &8.1  &9.2   &9.3   \\
          &   &      &      &     &     &      &      &     &     &      &     &      &      &      &     &      &      \\
1226$-$297&N  &17    &$-$   &12   & $-$ &12    &$-$   &10   &$-$  &6.8   &$-$  &5.3   &$-$   &6.8   &$-$  &5.3   &$-$   \\
          &S  &416   &12.8  &366  &14.4 &345   &13.2  &308  &14.8 &147   &13.7 &132   &15.6  &138   &14.2 &124   &16.1  \\
\end{tabular}
\end{table*}
\begin{table*}
\begin{tabular}{|l|l|r r r r|r r r r|r r r r|r r r r|}
\multicolumn{2}{c}{Source} & \multicolumn{4}{c}{22\,cm }
				& \multicolumn{4}{c}{18\,cm} & \multicolumn{4}{c}{6.47\,cm}& \multicolumn{4}{c}{6.08\,cm} \\
Name      &Cp&$S_I$ & m$_l$    &S$_p$&m$_h$&$S_I$ &  m$_l$   &S$_p$&m$_h$& $S_I$& m$_l$   &S$_p$ & m$_h$& $S_I$& m$_l$   &S$_p$ & m$_h$\\
          &    &      &      &     &     &      &      &     &     &      &     &      &      &      &     &      &      \\
1232$-$249&N  &1084  &9.6   &457  &13.0 &901   &10.1  &397  &12.8 &365   &9.4  &163   &11.7  &343   &8.9  &154   &11.6  \\
          &S  &878   &8.3   &405  &5.4  &722   &8.0   &357  &5.2  &297   &6.6  &169   &4.5   &280   &7.2  &158   &5.4   \\
1247$-$290&N  &416   &11.9  &290  &12.3 &335   &13.1  &234  &13.4 &121   &12.7 &88    &13.2  &115   &11.9 &80    &12.9  \\
          &S  &253   &11.9  &133  &8.6  &205   &11.5  &110  &8.7  &75    &5.9  &46    &4.7   &69    &8.3  &42    &6.8   \\
1257$-$230&N  &259   &5.2   &213  &4.1  &175   &5.8   &151  &4.4  &63    &3.8  &49    &3.7   &57    &4.3  &44    &3.9   \\
          &S  &514   &5.0   &456  &5.1  &398   &5.3   &373  &5.2  &167   &3.6  &147   &4.1   &158   &3.8  &139   &4.5   \\
1358$-$214&N  &211   &15.9  &58   &5.1  &163   &16.4  &45   &6.3  & $-$  & $-$ & $-$  & $-$  &$-$   & $-$ & $-$  & $-$  \\
          &S  &172   &13.7  &39   &11.3 &155   &11.6  &33.6 &12.2 & $-$  & $-$ & $-$  & $-$  &$-$   & $-$ & $-$  & $-$  \\
2035$-$203&E  &563   &13.8  &417  &13.1 &487   &13.5  &363  &13.8 &225   &13.6 &175   &14.7  &213   &14.8 &167   &15.5  \\
          &W  &154   &17.4  &21   &18.1 &126   &15.5  &17   &15.9 &43    &15.7 &7.2   &16.6  &39    &10.8 &6.7   &14.4  \\
          &   &      &      &     &     &      &      &     &     &      &     &      &      &      &     &      &      \\
2040$-$246&E  &247   &9.1   &105  &14.2 &203   &9.1   &86   &15.2 &82    &8.8  &37    &14.4  &76    &9.4  &35    &15.5  \\
          &W  &50    &6.5   &15   &6.2  &38    &4.3   &12   &5.6  &13    &$-$  &4.9   &$-$   &12    &$-$  &4.2   &$-$   \\
2042$-$293&N  &286   &9.6   &48   &9.1  &209   &10.4  &39   &8.7  &71    &4.8  &16    &6.6   &65    &6.3  &15    &6.6   \\
          &S  &128   &12.8  &23   &13.7 &88    &14.0  &17   &14.5 &32    &10.7 &7.4   &15.4  &26    &12.2 &6.8   &12.5  \\
2045$-$245&N  &383   &7.4   &180  &7.8  &303   &8.5   &146  &9.3  &96    &9.6  &52    &11.9  &88    &10.3 &48    &12.9  \\
          &S  &256   &15.1  &70   &17.7 &199   &14.2  &56   &18.0 &70    &11.79&20    &14.5  &62    &13.0 &18    &15.9  \\
2118$-$266&E  &277   &7.8   &78   &4.1  &212   &7.2   &65   &4.5  &84    &5.0  &28    &4.4   &77    &4.9  &26    &3.5   \\
          &W  &81    &8.5   &5    &12.0 &48    &9.7   &3.8  &13.7 &16    &$-$  &1.2   &$-$   &18    &$-$  &1.5   &$-$   \\
2132$-$236&N  &157   &10.8  &72   &9.1  &121   &11.2  &60   &9.5  &44    &11.2 &24    &9.9   &39    &9.7  &23    &10.0  \\
          &S  &177   &7.6   &122  &7.6  &136   &8.6   &100  &8.1  &48    &8.1  &36    &8.4   &43    &7.3  &33    &7.8   \\
          &   &      &      &     &     &      &      &     &     &      &     &      &      &      &     &      &      \\
2137$-$279&N  &170   &8.1   &46   &10.5 &132   &8.1   &37   &9.9  &46    &6.4  &14    &9.5   &42    &6.1  &13    &8.5   \\
          &S  &285   &11.4  &168  &12.6 &215   &13.5  &135  &13.8 &71    &13.1 &45    &13.8  &67    &14.3 &42    &14.4  \\
2213$-$283&E  &432   &14.0  &126  &21.9 &347   &13.3  &103  &21.7 &115   &12.2 &35    &20.1  &107   &11.7 &32    &19.4  \\
          &W  &328   &5.2   &166  &5.5  &253   &5.6   &131  &5.5  &73    &9.6  &41    &9.0   &67    &9.7  &37    &8.8   \\
2311$-$222&E  &816   &11.8  &577  &9.6  &666   &12.0  &505  &9.9  &278   &11.4 &205   &10.7  &262   &11.2 &194   &10.8  \\
          &W  &144   &7.0   &64   &4.0  &113   &5.8   &53   &3.9  &46    &3.8  &23    &3.1   &44    &2.4  &22    &2.9   \\
2325$-$213&N  &378   &13.7  &113  &9.9  &291   &13.5  &91   &10.7 &107   &11.3 &34    &9.5   &99    &11.3 &32    &9.5   \\
          &S  &608   &9.1   &136  &14.2 &461   &9.7   &109  &14.7 &159   &9.6  &63    &14.5  &146   &9.0  &40    &14.6  \\
2338$-$290&N  &166   &11.5  &58   &15.4 &128   &11.0  &47   &15.5 &56    &8.9  &21    &13.9  &49    &10.2 &21    &14.4  \\
          &S  &216   &15.0  &17   &16.3 &146   &15.6  &14   &15.2 &59    &13.8 &5.8   &15.0  &51    &12.7 &5.2   &17.0  \\
          &   &      &      &     &     &      &      &     &     &      &     &      &      &      &     &      &      \\
2348$-$290&N  &199   &13.5  &55   &13.6 &157   &13.2  &45   &12.9 &57    &10.0 &19    &8.8   &51    &9.7  &18    &8.6   \\
          &S  &299   &13.9  &69   &14.0 &235   &16.3  &57   &14.7 &83    &16.9 &22    &12.5  &77    &16.4 &21    &12.7  
\end{tabular}
\end{table*}

GCL91 suggested that X-ray emitting gas associated with poor
clusters of galaxies could  produce the observed depolarization in their
sample. Assuming similar parameters as those used by GCL91 for the 
depolarizing gas in an unresolved foreground Faraday screen (Burn 1966),
and using the median values of size and redshift 
for our sources, we expect the  depolarization between $\lambda$20 and 
$\lambda$6\,cm in the observed frame to be close to about 1 for galaxies
and about 0.8 for quasars. However, there are significant uncertainties
in the assumed parameters of GCL91, and the Faraday screens may also 
have partially resolved structures. We hope to place better constraints on 
these values after making observations over a longer wavelength range. 
Significantly stronger depolarization will be seen at longer wavelengths, say
between $\lambda$20\,cm and either $\lambda$49\,cm or $\lambda$90\,cm. 
Since long-wavelength measurements are still being planned using the GMRT,
we discuss in this paper any possible dependence of 
depolarization on linear size and lobe separation ratio 
using the present data. We also discuss the depolarization asymmetry of 
the oppositely directed lobes for both radio galaxies and quasars and comment
on whether this is consistent with the unified scheme. 
In the following sections, we first present the data for the entire 
sample and then focus on the strong-lobe sample where the 
errors are smaller and 
the effect of poor signal to noise ratio will be minimum. 

\begin{table} \caption {Derived parameters of the lobes and hotspots}
\begin{tabular}{ l l r r r r r }
Source   &   L & $r_{\theta}$  & DP$_l$   & $\sigma_{_{DP_l}}$ &   DP$_h$   & $\sigma_{_{DP_h}}$ \\
Name   & kpc &               &          &                    &            &                    \\
       &   &  & &         &   &       \\
       &       &        &        &         &        &         \\
0017$-$207N & 348 & 0.93 & 0.98 & 0.04 & 0.91 & 0.03\\
0017$-$207S & 374 & 1.08 & 1.24 & 0.08 & 1.84 & 0.19\\
0058$-$229N & 201 & 0.66 & 0.68 & 0.05 & 0.80 & 0.04\\
0058$-$229S & 305 & 1.52 & 1.29 & 0.09 & 1.27 & 0.07\\
0133$-$266N & 189 & 0.71 & 0.94 & 0.10 & 1.11 & 0.13\\
0133$-$266S & 266 & 1.41 & 1.08 & 0.09 & 0.98 & 0.07\\
       &         &     &         &         &        &         \\
0137$-$263N & 323 & 0.89 & 0.92 & 0.03 & 0.59 & 0.04\\
0137$-$263S & 363 & 1.13 & 1.09 & 0.06 & 0.99 & 0.05\\
0148$-$297N & 509 & 1.11 & 1.04 & 0.03 & 0.93 & 0.02\\
0148$-$297S & 458 & 0.90 & 1.04 & 0.03 & 1.02 & 0.02\\
0325$-$260N & 241 & 1.17 & 1.63 & 0.19 & 1.93 & 0.39\\
0325$-$260S & 206 & 0.85 & 1.21 & 0.10 & 1.27 & 0.13\\
       &         &     &         &         &        &         \\
0346$-$297S & 383 & 0.70 & 1.14 & 0.04 & 1.22 & 0.07\\
0428$-$281E & 352 & 1.32 & 1.24 & 0.07 & 1.02 & 0.08\\
0428$-$281W & 266 & 0.76 & 0.96 & 0.04 & 0.87 & 0.06\\
0437$-$244N & 415 & 0.64 & 1.02 & 0.07 & 1.13 & 0.06\\
0437$-$244S & 645 & 1.55 & 1.83 & 0.20 & 1.82 & 0.28\\
       &         &     &         &         &        &         \\
0454$-$220N & 239 & 0.54 & 0.65 & 0.03 & 0.60 & 0.04\\
0454$-$220S & 445 & 1.86 & 0.98 & 0.04 & 0.95 & 0.06\\
0551$-$226N & 258 & 1.43 & 0.82 & 0.05 & 0.67 & 0.05\\
0551$-$226S & 180 & 0.70 & 0.87 & 0.05 & 0.96 & 0.07\\
0937$-$250N & 270 & 0.80 & 0.82 & 0.06 & 1.74 & 0.32\\
0937$-$250S & 339 & 1.25 & 1.07 & 0.09 & 0.98 & 0.13\\
       &         &     &         &         &        &         \\
0938$-$205S & 328 & 2.79 & 1.06 & 0.05 & 1.07 & 0.17\\
0947$-$249N & 242 & 0.69 & 0.64 & 0.02 & 0.92 & 0.03\\
0947$-$249S & 353 & 1.46 & 1.01 & 0.07 & 0.86 & 0.08\\
0955$-$283E & 375 & 0.90 & 1.02 & 0.05 & 1.08 & 0.06\\
0955$-$283W & 417 & 1.11 & 1.00 & 0.05 & 0.99 & 0.05\\
       &         &     &         &         &        &         \\
1022$-$250E & 251 & 2.15 & 0.88 & 0.11 & 1.32 & 0.07\\
1022$-$250W & 116 & 0.46 & 1.03 & 0.04 & 1.28 & 0.19\\
1023$-$226N & 214 & 0.75 & 1.24 & 0.13 & 1.33 & 0.17\\
1023$-$226S & 285 & 1.34 & 1.10 & 0.24 & 1.03 & 0.33\\
1025$-$229N & 568 & 1.05 & 1.05 & 0.08 & 1.03 & 0.12\\
       &         &     &         &         &        &         \\
1026$-$202N & 241 & 1.11 & 1.19 & 0.06 & 1.03 & 0.07\\
1026$-$202S & 217 & 0.90 & 0.88 & 0.06 & 0.86 & 0.07\\
1029$-$233E & 299 & 0.97 & 0.99 & 0.03 & 0.82 & 0.03\\
1029$-$233W & 310 & 1.03 & 1.03 & 0.07 & 1.01 & 0.12\\
1052$-$272N & 197 & 0.35 & 0.74 & 0.05 & 0.66 & 0.05\\
1052$-$272S & 564 & 2.86 & 1.18 & 0.08 & 0.94 & 0.10\\
       &         &     &         &         &        &         \\
1107$-$218E & 269 & 1.02 & 0.87 & 0.09 & 1.10 & 0.16\\
1107$-$218W & 264 & 0.98 & 0.72 & 0.04 & 0.86 & 0.04\\
1107$-$227N & 274 & 0.87 & 1.03 & 0.06 & 0.45 & 0.05\\
1107$-$227S & 316 & 1.15 & 1.31 & 0.06 & 1.78 & 0.14\\
1126$-$290N & 401 & 1.24 & 0.85 & 0.04 & 0.59 & 0.12\\
1126$-$290S & 325 & 0.81 & 0.99 & 0.04 & 1.04 & 0.05\\
       &         &     &         &         &        &         \\
1224$-$208N & 156 & 0.43 & 0.85 & 0.03 & 0.92 & 0.04\\
1224$-$208S & 363 & 2.32 & 1.25 & 0.11 & 1.20 & 0.14\\
1226$-$297S & 309 & 1.07 & 0.90 & 0.03 & 0.90 & 0.03\\
       &         &     &         &         &        &         \\

\end{tabular}
\end{table}

\begin{table}

\begin{tabular}{ l l r r r r r }
   &                 &       &       &         &         &       \\
   &                 &       &       &         &         &       \\
Source   &   L & $r_{\theta}$  & DP$_l$   & $\sigma_{_{DP_l}}$ &   DP$_h$   & $\sigma_{_{DP_h}}$ \\
Name   & kpc &             &          &                    &            &                    \\
  &                  &  & &         &  &       \\

            &      &       &      &      &      &     \\
1232$-$249N & 356 & 1.17 & 1.08 & 0.05 & 0.98 & 0.08\\
1232$-$249S & 304 & 0.85 & 1.16 & 0.07 & 1.05 & 0.08\\
1247$-$290N & 195 & 0.66 & 1.00 & 0.04 & 0.97 & 0.04\\
1247$-$290S & 297 & 1.52 & 1.43 & 0.08 & 1.22 & 0.09\\
1257$-$230N & 217 & 0.88 & 1.22 & 0.12 & 1.21 & 0.15\\
1257$-$230S & 245 & 1.13 & 1.32 & 0.13 & 1.42 & 0.16\\
            &       &      &      &      &      &     \\
2035$-$203E & 279 & 1.13 & 0.93 & 0.03 & 0.85 & 0.03\\
2035$-$203W & 246 & 0.88 & 1.60 & 0.09 & 1.26 & 0.10\\
2040$-$236E & 153 & 0.43 & 0.97 & 0.05 & 0.97 & 0.05\\
2042$-$293N & 383 & 1.40 & 1.52 & 0.11 & 2.70 & 0.64\\
2042$-$293S & 274 & 0.71 & 1.05 & 0.07 & 1.37 & 0.14\\
       &           &     &         &         &        &         \\
2045$-$245N & 269 & 0.78 & 0.73 & 0.04 & 0.70 & 0.04\\
2045$-$245S & 343 & 1.27 & 1.16 & 0.05 & 1.14 & 0.04\\
2118$-$266E & 247 & 0.88 & 1.61 & 0.15 & 1.20 & 0.19\\
2132$-$236N & 275 & 1.33 & 1.11 & 0.07 & 0.92 & 0.05\\
2132$-$236S & 207 & 0.75 & 1.05 & 0.07 & 0.98 & 0.07\\
       &           &     &         &         &        &         \\
2137$-$279N & 250 & 1.30 & 1.33 & 0.11 & 1.16 & 0.15\\
2137$-$279S & 193 & 0.77 & 0.80 & 0.03 & 0.91 & 0.03\\
2213$-$283E & 469 & 2.86 & 1.20 & 0.04 & 1.15 & 0.03\\
2213$-$283W & 164 & 0.35 & 0.54 & 0.04 & 0.62 & 0.04\\
2311$-$222E & 281 & 0.84 & 1.05 & 0.04 & 0.94 & 0.04\\
2311$-$222W & 335 & 1.19 & 2.95 & 0.47 & 1.33 & 0.21\\
       &           &     &         &         &        &         \\
2325$-$213N & 355 & 1.27 & 1.21 & 0.05 & 1.11 & 0.06\\
2325$-$213S & 278 & 0.78 & 1.01 & 0.05 & 0.97 & 0.04\\
2338$-$290N & 275 & 0.98 & 1.13 & 0.07 & 1.07 & 0.04\\
2338$-$290S & 282 & 1.03 & 1.18 & 0.07 & 1.02 & 0.12\\
2348$-$235N & 285 & 1.00 & 1.40 & 0.07 & 1.32 & 0.11\\
2348$-$235S & 286 & 1.00 & 0.84 & 0.03 & 1.09 & 0.05\\
\end{tabular}
\end{table}

\begin{figure}[t]
\vbox{
\hbox{
\hspace{-0.8in}
\psfig{figure=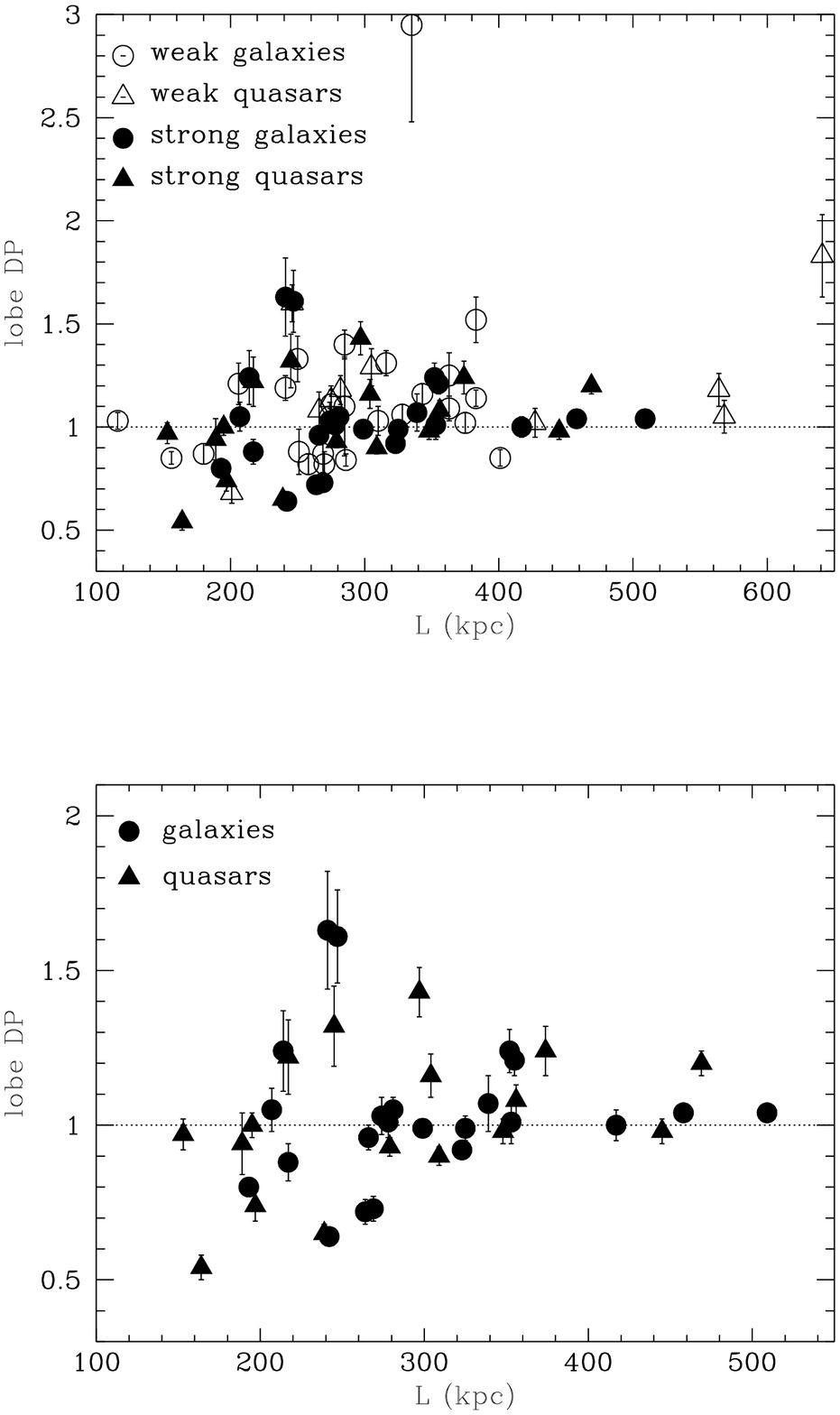,height=6.8in,width=5.0in}
}
\vspace{0.2in}
\caption{ Depolarization of each lobe against its linear 
separation from the core or optical position for the entire sample (upper
panel)  and for the strong lobes (lower panel). }
}
\end{figure}

\begin{figure}[t]
\vbox{
\hbox{
\hspace{-0.7in}
\psfig{figure=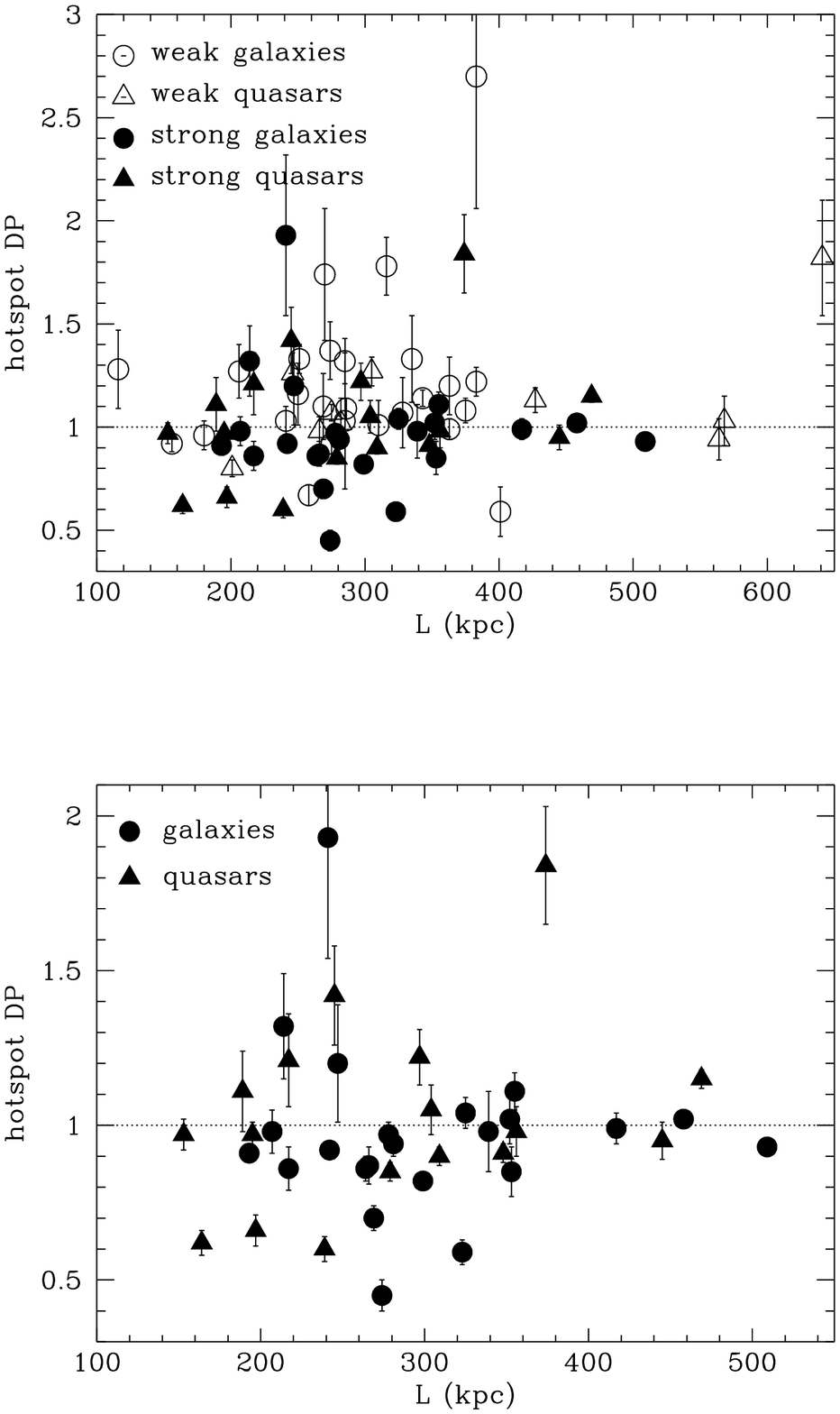,height=6.8in,width=5.0in}
}
\vspace{0.2in}
\caption{ Depolarization of the hotspots against their linear 
separations from the core or optical position for the entire sample
(upper panel) and for the sample of strong lobes (lower panel). }
}
\end{figure}

\subsection{Linear separation of the lobe and depolarization}
We present the plot of the linear separation 
of each lobe from the core or optical position against the 
depolarization parameter, DP=m$_{20}$/m$_6$,  of the entire lobe in
Figure 2, and against the corresponding DP values of the hotspots in 
Figure 3. The upper panel  in each figure shows the plot for the 
entire data while the bottom one shows only for the sample of strong lobes 
which have been defined to have S$_{peak}$(6\,cm) $> 25$ mJy. These
figures show that evidence of depolarization is seen more frequently
in the sources of smaller linear size. Considering the entire sample,
20 of the 43 lobes with a linear separation $L <$
300 kpc have DP $<$ 1, while for those with $L >$
300 kpc, only 7 of 33 have DP $<$ 1. A Kolmogorov-Smirnov test shows that
the DP distributions for lobes with L $<$ 300 and $>$ 300 kpc are different
at a significance level of about 99 per cent.
A similar but weaker
trend is seen for the hotspots.
Concentrating only on those which show 
significant depolarization, say DP$<$ 0.9, there are 17 lobes, 15
of which have L$<$ 300 kpc, and  15 hotspots, 11 of which have L$<$ 300 kpc.

These trends for the sample of strong lobes are seen clearly 
in the Figures.
Above a linear separation, {\it L}, of 300 kpc, there is little 
evidence of depolarization, while lobes and hotspots  below this 
value often have DP significantly less than 1. 
Here 13 of the 24 lobes with  {\it L}$<$ 300 kpc have 
DP $<$ 1.0, while this is true for only 5 of 16 with {\it L}$>$300 kpc.
All the 8 lobes with DP $<$ 0.9 have {\it L}$<$ 300 kpc. A similar trend 
is also seen while considering the values for the hotspots
where 10 of 24 with {\it L}$<$300 kpc have DP$<$ 0.9  while this
is true for only 2 of the 16 lobes with {\it L}$>$300 kpc.

GCL91 reported that the depolarization parameter on the counter-jet side,
DP$_{cj}$, decreases 
with increasing distance from the core, while for the jet side, DP$_{j}$,
no such trend is seen. For a sample of 12 galaxies, observed by 
Pedelty et al. (1989) with the VLA at $\lambda$20 and 6\,cm, DP 
is about 0.3 for the smaller sources with linear separations $L$ within 
about  100 kpc, and 
increases to about 1 for a separation of about 300 kpc which is 
amongst their largest objects. Our source sizes are about a factor of two
larger than those of Pedelty et al. The smallest separation in our sample 
of strong lobes is about 150 kpc 
and has a DP of about 0.54 $\pm$ 0.04. 
For $L >$  300 kpc, almost all sources show no evidence of strong 
depolarization. However, a number of sources appear to have DP $>$ 1. We
have examined the data for both strong and weak sources, and also for sources
observed on different days and find no reasonable cause for any systematic error. 
A similar 
trend for dependence of DP on $L$ was also reported for relatively nearby radio galaxies by 
J\"{a}gers (1986) and Strom \& J\"{a}gers (1988) from observations 
with the WSRT at $\lambda$20 and 49\,cm. They find
DP$^{49}_{21}$ to be about 0.6 at $L$ of about 250 kpc, which is 
consistent with our DP$^{20}_{6}$ values at this separation. 

Considering the galaxies and quasars separately, there is no 
striking difference in their depolarization values. 
This is  partly because we have selected large sources.
The observed 
depolarization is likely to be affected by intrinsic asymmetries 
in addition to any path-length differences between quasars and radio galaxies. 
We examine the  depolarization asymmetry between the oppositely 
directed lobes for radio galaxies and quasars in Section 4.4.

\begin{figure}[t]
\vbox{
\hbox{
\hspace{-0.8in}
\psfig{figure=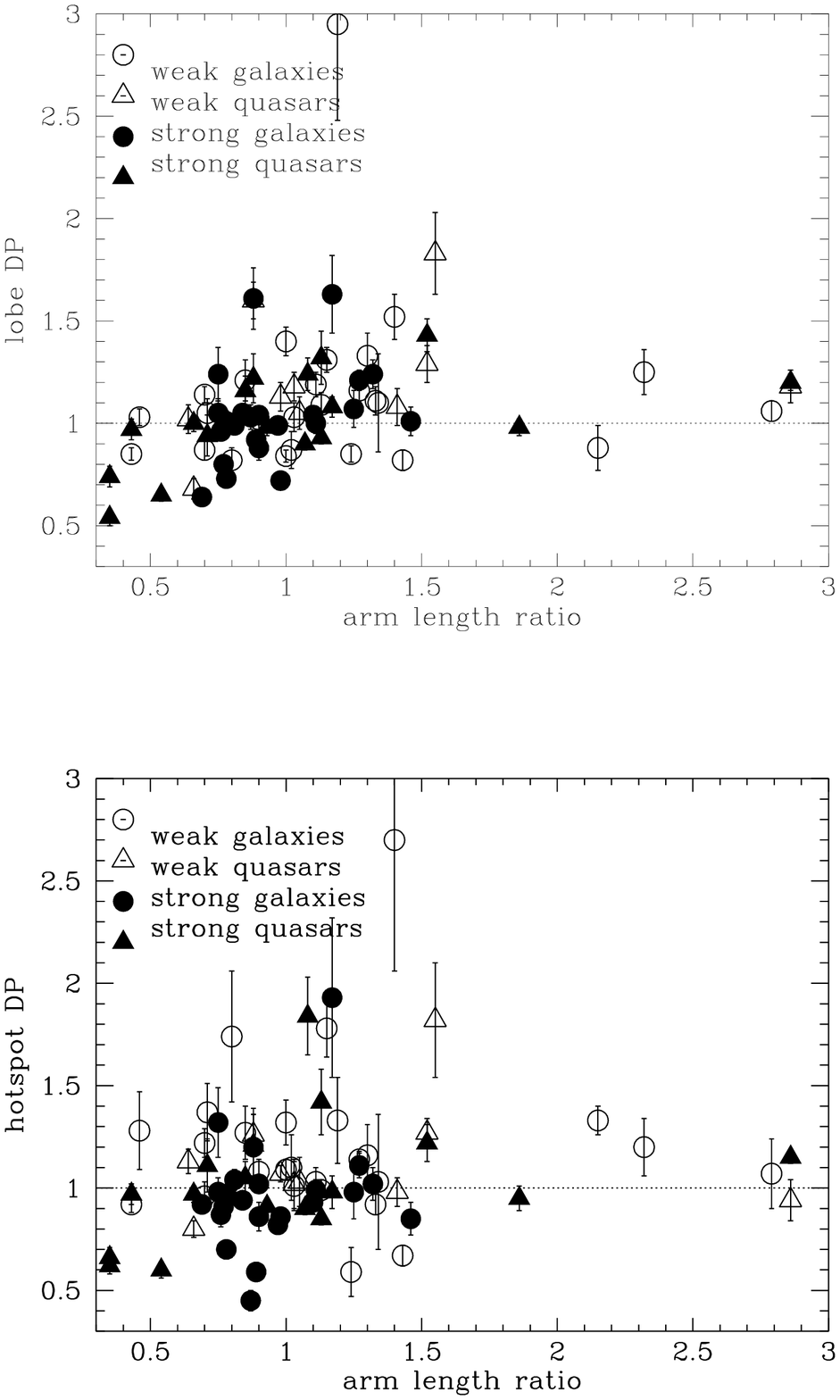,height=6.8in,width=5.0in}
}
}
\caption{ The depolarization of each lobe (upper panel) and hotspot (lower
panel) against the arm-length ratio for the entire sample with filled points 
denoting the sample of strong lobes.}
\end{figure}

\subsection{Lobe separation ratio and depolarization}
For a sample of 3CR radio galaxies, 
McCarthy, van Breugel \& Kapahi (1991) showed that there is more 
emission-line gas on the side of the lobe which is closer to the 
nucleus, suggesting an environmental origin for the observed 
arm-length asymmetries rather than being due to light travel time 
across the source axes. 
For the sample of galaxies observed by Pedelty et al. (1989), the 
depolarization is almost always stronger for the lobe closer 
to the nucleus. In 5 of their 8 sources that have extended 
emission-line gas, there is a spatial correlation between 
depolarization and emission-line gas, while for the remaining three, 
the gas is more centrally located and the sources are not strongly 
depolarized.  Our sample of sources has not been chosen on the 
basis of their emission-line properties which are being presently
investigated.
 
In Figure 4 (upper panel) we plot for each lobe the ratio, 
r$_\theta$ (arm-length ratio), of its separation from the nucleus 
to that of the opposite lobe, against the depolarization parameter 
of the lobe. As we are plotting the DP for each lobe or hotspot with
significant polarized flux density at $\lambda$20 and 6 cm, r$_\theta$
is $<$ 1 for the nearer lobe or hotspot and $>$ 1 for the farther one.
There is a trend for the lobes with DP$<$1 to have r$_\theta$ $<$1,
i.e. they are
closer to the nucleus compared to the lobes on the opposite side.
For the entire sample of 76 lobes, 
20 of 39 with r$_\theta$ $<$1 have DP $<$ 1, while of those
with r$_\theta$ $>$1 this is true for only 7 of the 37 lobes. 
A Kolmogorov-Smirnov test shows that the DP distributions for sources
with r$_\theta$ $<$ 1 is different from those with r$_\theta$ $>$ 1
at a significance level of $>$ 99 per cent. 
Concentrating on the sample of strong lobes, DP $<$ 1 for  
15 out of 25 lobes with r$_\theta$ $<$1, 
but for only 3 out of 15 with r$_\theta$ $>$1. 
Among the eight lobes with significant depolarization, 
DP $<$ 0.90, all are on the nearer side with
r$_\theta$ $<$1. A similar trend is also seen for the hotspots.
There are 12 hotspots with DP $<$ 0.9, 10 of which have r$_\theta$ $<$1.

The tendency for the shorter arm to be more depolarized 
can also be seen clearly if the  
arm-length ratio, defined to be the ratio of the length of the 
short arm to the length of the longer arm, is  plotted against 
the depolarization of the short arm to the depolarization of the 
longer arm (Figure 5).   It is seen that in 28 
of the 35 sources, where the depolarization of the both the lobes
have been determined,  the nearer lobe is more depolarized than the farther 
one. There is a weak trend for the  most asymmetric 
sources to have a lower  depolarization ratio, similar to what 
was observed by Pedelty et al. (1989).  This is seen more clearly
in Figure 5 when one considers the galaxies and quasars with strong lobes.
The dependence of depolarization ratio on arm-length ratio was studied by 
GCL91, but they did not find any such trend.  Their 
sample consists largely of quasars with known radio jets and are of smaller 
angular size compared to our sources.  In a sample of quasars selected to have 
known radio jets, the objects are likely to be close to the line
of sight and effects of orientation should be dominant. The counter-jet side which 
exhibits significantly stronger depolarization is often farther
from the nucleus. This is possibly due to intrinsic non-collinearities which
may also appear amplified if the sources are at small angles to the line of sight.
In our sample the sources are of large angular and linear size, reasonably
collinear and the observed trend is possibly due to an asymmetric environment
as well as effects of orientation. Comparison of the depolarization information
with narrow-band images, and detection of radio jets in these objects should
help clarify the situation.

\begin{figure}[t]
\vbox{
\hbox{
\hspace{-0.8in}
\psfig{figure=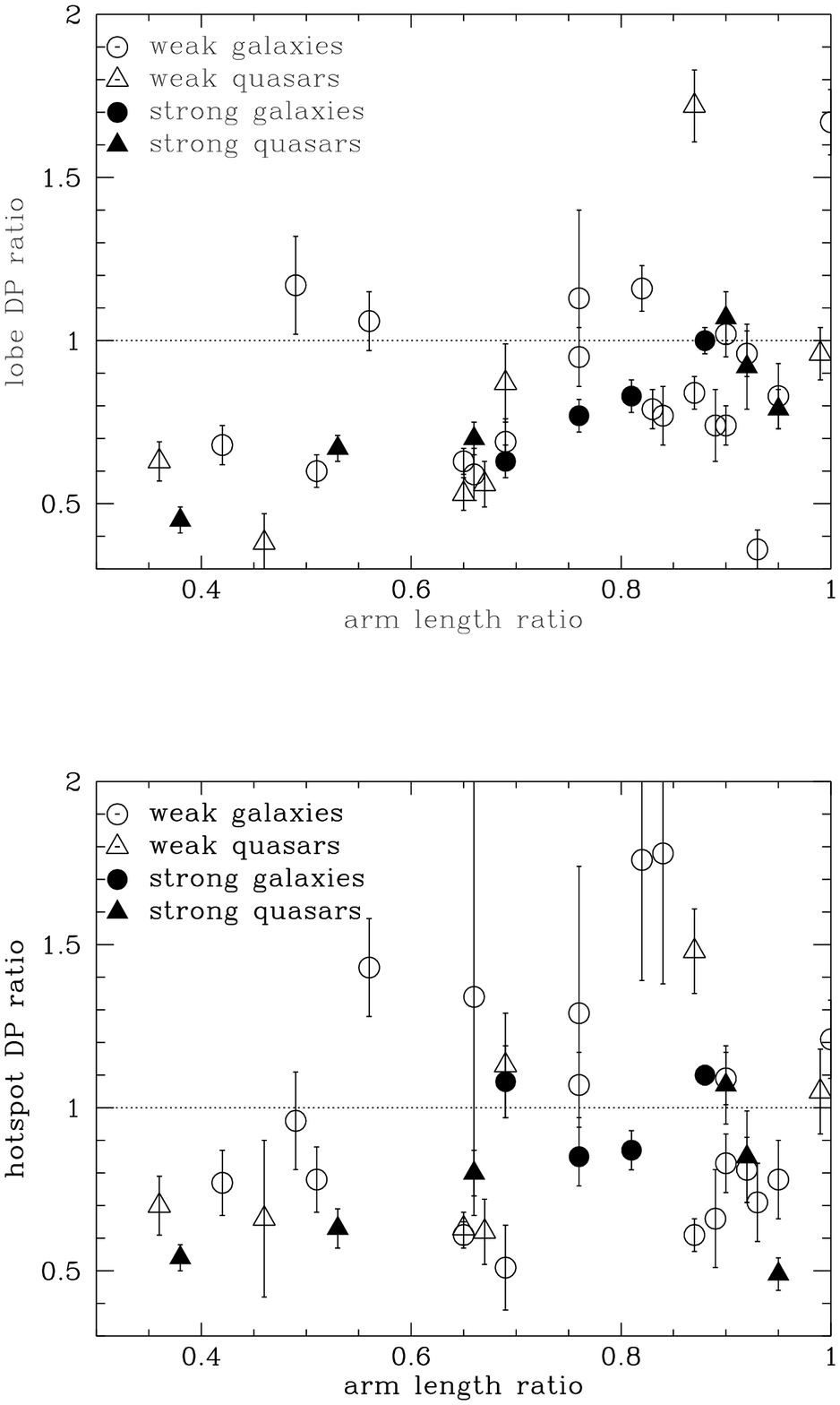,height=6.8in,width=5.0in}
\vspace{0.3in}
}
}
\caption{ The depolarization of the nearer lobe (upper panel) or hotspot 
(lower panel) to the farther 
one for each source against the arm-length ratio, which is
now defined to be always $\leq$1.  The sample of strong lobes is denoted
by the filled points.}
\end{figure}

\subsection{Flux density and brightness ratio of the lobes}
However, an examination of the brightness asymmetry of the lobes and its
relationship to the arm-length asymmetry might also provide us with valuable 
insights. For example, if the source and the environment are intrinsically
symmetric, the approaching
component would be farther from the nucleus and brighter due to relativistic
enhancement of the hot-spot flux density, provided the effects of evolution
of individual components with age
over the length scales of our sources are not signficant (Ryle \& Longair 1967;
Swarup \& Banhatti 1981). On the other hand, if the source is evolving in an
asymmetric environment, there will be a greater dissipation of energy on the
side with the higher density which will also be closer to the nucleus
(Eilek \& Shore 1989; Gopal-krishna \& Wiita 1991). Figure 6 shows the 
arm-length ratio, defined to be $\leq$1, plotted against the peak brightness 
ratio, which is the ratio of peak brightness of the nearer lobe to the farther one. 
There is a clear trend for the nearer component to be brighter, so that the ratio
is $>$1 in 29 of the 41 objects in the sample. This is particularly true for the
more asymmetric sources, say r$_\theta <$0.8, where 18 of the 22 sources have the 
brighter component closer to the nucleus. A similar trend is also seen while 
considering the total flux density ratio of the lobes. The sample of 3CR sources
studied by McCarthy et al. (1991) shows a similar but weaker trend. This trend 
in our sample suggests that an asymmetric environment on opposite sides of 
the nucleus affects the flux density and arm-length ratios of the lobes, in 
addition to the observed depolarization asymmetry of the oppositely-directed lobes.

\begin{figure}[t]
\vspace{-1.0in}
\psfig{figure=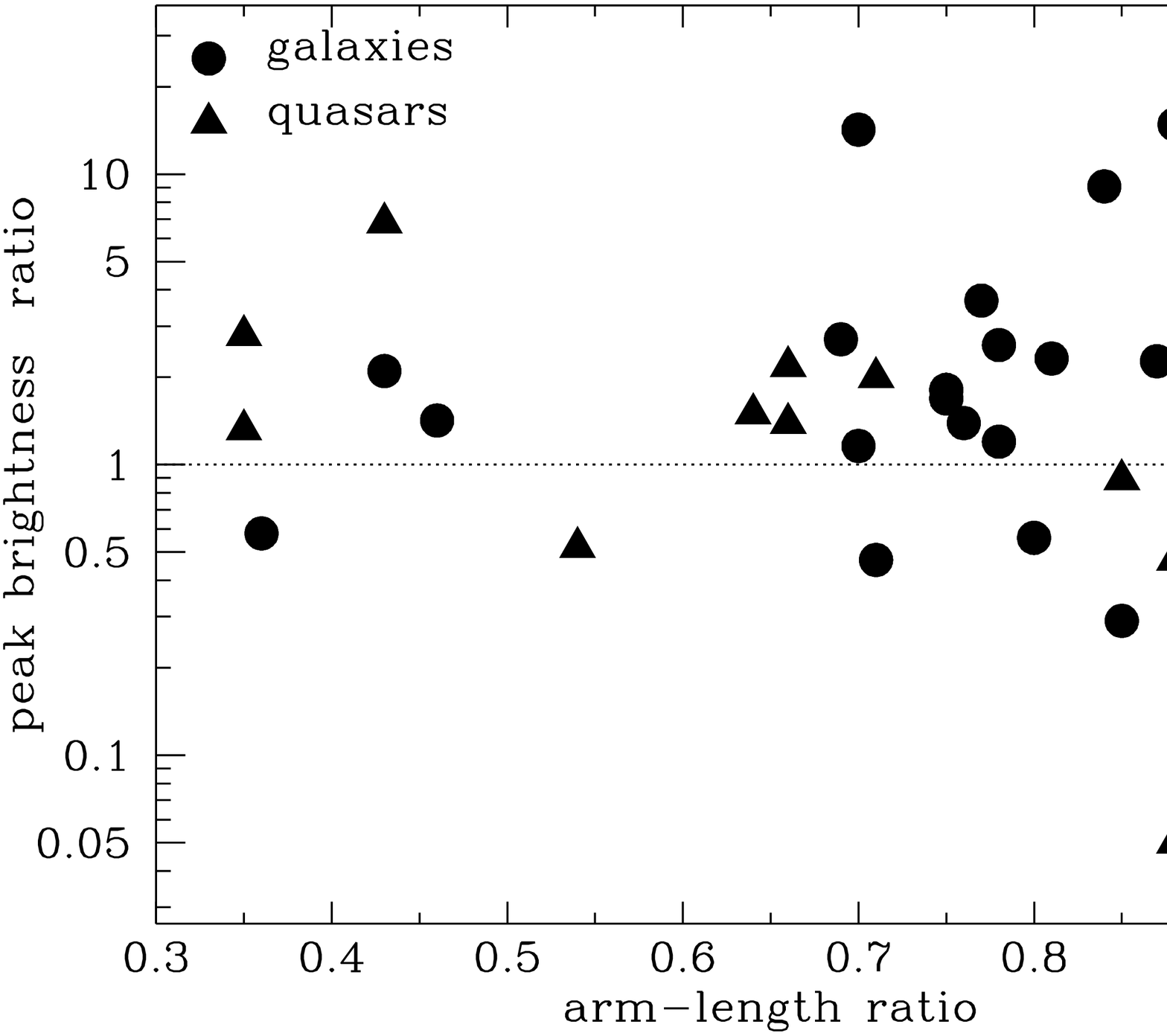,height=3.5in,width=2.6in}
\caption{Ratio of peak brightness at $\lambda 20$\,cm of the nearer lobe to the 
farther one against the corresponding arm-length ratio. }
\end{figure}

\begin{figure*}[t]
\vbox{
\vspace{-1.8in}
\psfig{figure=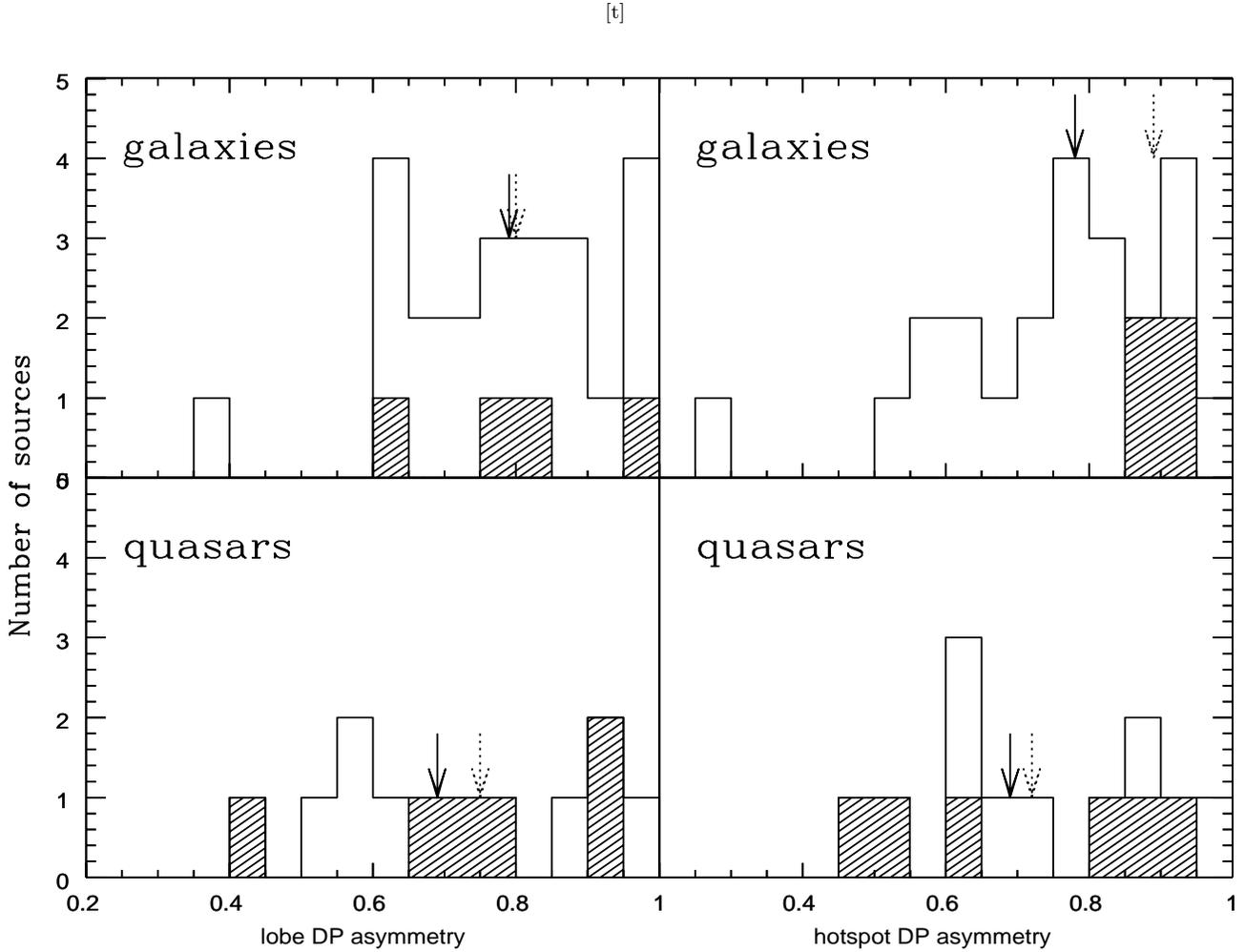,height=7.0in,width=5.6in}
\caption{The distributions of the ratio of depolarization of the lobes (left panel) and
hotspots (right panel) on opposite sides of the parent galaxy. The ratio is defined to be 
$\leq$1, and is shown shaded for the sample of strong lobes. The solid and dotted arrows
 show the median values for the entire sample and the sample of strong lobes respectively.}
}
\end{figure*}

\subsection{Depolarization Asymmetry} 
In the unified scheme for radio galaxies  and quasars, the 
quasars are inclined at smaller angles to the line of sight. 
If the depolarization is caused by a magnetioionic halo 
associated with the parent optical object or a cluster of galaxies
associated with it, the pathlength difference for the radiation from
the two oppositely-directed lobes will
be larger for quasars compared to the galaxies. Hence the quasars should
exhibit a higher degree of depolarization asymmetry of the lobes or
hotspots compared to the galaxies. We can attempt to infer an average
statistical angle of inclination to the line of sight for our sample
of galaxies and quasars using the fraction of emission from the core
as a statistical indicator of orientation to the line of sight. 
All the quasars in our sample have detected radio cores, 
with the median fraction
of emission from the core at an emitted frequency of 8\,GHz being about 12\%.
Most of the galaxies do not have detected
cores, and the upper limit is generally less than about 0.1\%.
Assuming that the intrinsic fraction of emission from the nucleus
is similar to that of 3CR sources (cf. Saikia \& Kulkarni 1994), the 
core strength of galaxies and quasars are consistent with angles of 
inclination of about 65$^\circ$ and 30$^\circ$  respectively.
The expected depolarization asymmetry between $\lambda$20 and 6\,cm
for our sources using the typical parameters listed by GCL91 are close to about
1 and 0.8 for the galaxies and quasars respectively. However, our sources
are much larger than those in GCL91, and we need to determine more
reliably the parameters for the environment on these scales using X-ray and
optical observations along with long-wavelength radio polarization measurements.

Since we do not 
detect jets in almost all but one of our sources we examine the DP
asymmetry for both radio galaxies and quasars, bearing in mind that
the trend could be diluted by intrinsic asymmetries in the environment.
In Figure 7, we present the depolarization asymmetry for both radio galaxies and
quasars using the depolarization values for both the lobes and hotpots. 
We define
the depolarization asymmetry as the ratio of the depolarization of one lobe
to the other such that the ratio is always less than 1. There is
a marginal trend for the quasars to be more asymmetric, consistent with 
the sense expected in the unified scheme. The median values for the entire
sample of galaxies and quasars are about 0.78 and 0.67 respectively for the
lobes as well as the hotspots. Considering the lobes and hotspots from the
sample of strong sources, the median values of DP asymmetry of the lobes are 
0.80 and 0.75 for the galaxies and quasars respectively, while the 
corresponding values for the hotspots are 0.89 and 0.72 respectively. Since
the environments are likely to be asymmetric, as discussed earlier, and the
sources are large, the effects of orientation are seen as a marginal trend.
\subsection{Laing-Garrington effect}
In the Laing-Garrington effect the counter-jet side depolarises more 
rapidly due to the extra pathlength through the magnetoionic medium. 
In our sample we detect a radio jet in only one object, namely the quasar
0454$-$220.  The depolarization value for the lobe and the hotspot on
the counter-jet side is 0.65$\pm$ 0.03 and 0.60$\pm$ 0.04
and for  the jet side these values are
0.98$\pm$ 0.04 and 0.95$\pm$ 0.04 respectively. This is clearly 
consistent with the Laing-Garrington effect. 
The arm-length
ratio, r$_\theta$ defined to be $<$1, is 0.53 with the counter-jet lobe 
being the nearer one, while the DP ratio of this
lobe to the farther one is 0.67 $\pm$ 0.04 and
0.63 $\pm$ 0.06 for the hotspot. The component on the jet side is brighter,
possibly due to mild relativistic beaming of the hotspot. In this case, 
since we can unambiguously
identify the approaching component assuming the jet asymmetry to be
due to relativistic beaming, the tendency for r$_\theta$ and DP to be
$<$1, is largely due to the extra pathlength of the receding lobe.
It is important to try and detect radio jets in a larger number of these
objects to assess the relative importance of orientation and environmental
effects.

\section{Concluding remarks}
We have investigated the effects of environment and orientation on the 
observed depolarization properties of a sample of high-luminosity radio galaxies
and quasars between $\lambda$ 20 and 6\,cm. We find that significant 
depolarization is usually seen in the lobes which are within about 300 
kpc of the parent galaxy. 
Among the 17 lobes in the entire sample which show significant depolarization with DP$<$0.9, 
15 are within 300 kpc from the parent galaxy. Comparing the depolarization on opposite sides
of the source, we find that the side which is closer to the parent galaxy 
shows significant depolarization. Of these 17 lobes where significant depolarization is seen,
13 of them are closer to the parent galaxy compared to the lobe on the opposite side. 
The ratio of the depolarization parameter of the nearer lobe to the farther one is $<$1 for 28 of the
35 sources in the entire sample with reliable polarization information for both lobes.
The nearer component is also brighter in 26 of these 35 objects, suggesting that the 
nearer component is advancing outwards through a denser environment which is responsible
for stronger depolarization and greater dissipation of energy.
The depolarization asymmetry of the lobes on opposite sides for galaxies and quasars shows
that the latter  are marginally more asymmetric, consistent with the trend expected in 
the unified scheme. In our sample of sources the polarization properties of the lobes appear
to be due to an asymmetric environment on opposite sides of the parent optical object, as
well as possibly due to different orientations of the sources to the line of sight. Detailed  
information on
the distribution of the depolarizing medium from optical and X-ray observations, as well as 
detection of radio jets in these objects  to identify the approaching and receding components,
should enable us the clarify better the relative importance of the effects of 
an asymmetric environment and orientation. 

\subsection*{Acknowledgements}
The National Radio Astronomy Observatory  is a 
facility of the National Science Foundation operated under co-operative
agreement by Associated Universities Inc. We thank 
the staff of the Very Large Array for the observations, and Gopal-Krishna
and A. Pramesh Rao for their comments on the manuscript.

{}

\end{document}